# Structural Properties of the Lattice Heavy Quark Effective Theory

Jeffrey E. Mandula[a] and Michael C. Ogilvie[b]

[a]Department of Energy, Division of High Energy Physics
Washington, DC 20585

[b]Department of Physics, Washington University
St. Louis, MO 63130

We discuss two related aspects of the lattice version of the heavy quark effective theory (HQET). They are the effects of heavy quark modes with large momenta, near the boundary of the Brillouin zone, and the renormalization of the lattice HQET. We argue that even though large momentum modes are present, their contributions to heavy-light bound states and perturbative loop integrals are dynamically suppressed and vanish in the continuum limit. We also discuss a new feature of the renormalization of the lattice HQET not present in the continuum theory, namely that the classical velocity is finitely renormalized.

## 1. INTRODUCTION

The Isgur-Wise form of the heavy quark effective theory (HQET) is based on the separation of mass scales and on the physical idea that the motion of a very heavy object is only slightly perturbed by its interaction with much lighter objects. The idea of the Isgur-Wise limit is that a heavy quark's momentum can be usefully decomposed as $P_\mu = M v_\mu + p_\mu$, where $v$ is a constant but arbitrary four-velocity normalized to $v^2 = 1$ .[1,2] In the $M \to \infty$ limit, the fermion propagator simplifies to

$$\frac{1}{\gamma \cdot P - M} \to \frac{1 + \gamma \cdot v}{2 v \cdot p} \qquad (1)$$

The power of this limit is that the large momentum $Mv$ disappears, leaving behind only terms involving $v$ and $p$, which are $O(1)$ in the limit. Another important simplification is that the Dirac matrix structure of the propagator becomes trivial. This is responsible for the spin symmetry that appears in the Isgur-Wise limit. While the same physics lies behind both the static quark analysis[3] widely used in lattice simulations and the Isgur-Wise form of the HQET, the fact that $v$ can be non-zero in the Isgur-Wise form allows the calculation of processes with large momentum transfer, on the order of the heavy quark mass, such as the Isgur-Wise universal form factor describing the semileptonic decay of one particle containing a single heavy quark into another.

We have given earlier a lattice implementation of the HQET.[4] It shares with the continuum version the possibility that $v$ can take any value, and yet the momenta that enter simulations or perturbative loops remain small, as is appropriate for lattice analysis.

This lattice heavy quark effective theory (LHQET) has some unfamiliar features which we will discuss in this talk. One is the that there are modes of the heavy quark field that grow rather than decay with Euclidean time. A related property of the LHQET is that the heavy fermion modes are doubled. We will argue that the potential unphysical effects of these properties are absent, and that the same physical mechanism operates in both cases.

The other feature of the lattice HQET we will address here is that the classical velocity $v$ is (finitely) renormalized. We will show how this new renormalization arises from the lattice discretization of the reduced Dirac equation for the heavy quark propagator. We will give a one-loop perturbative estimate of its magnitude, and also compute it in a lattice simulation. The estimates are in fairly good quantitative agreement.

## 2. THE LATTICE HQET

The heavy quark effective theory is efficiently formulated[5] by factoring the $M \to \infty$ singular behavior from the field of a heavy quark, leaving a reduced operator $h^{(v)}$.

$$h^{(v)}(x) = e^{-iMv\cdot x} \frac{1+\gamma\cdot v}{2} \psi(x) \quad (2)$$

There is an independent field for each classical velocity. The Lagrangian for $h^{(v)}$ is

$$\mathcal{L}^{(v)} = \overline{h}^{(v)}(x)\, iv\cdot D\, h^{(v)}(x) \quad (3)$$

and its propagator satisfies the reduced Dirac equation

$$-iv\cdot D\, S = \delta \quad (4)$$

Here $D$ is the covariant derivative. The discretization by which we defined the lattice HQET is

$$\begin{aligned}
&v_0[\, U(x,x+\hat{t})\, S(x+\hat{t},y) - S(x,y)\,] \\
&+\sum_{\mu=1}^{3} \frac{-iv_\mu}{2} [U(x,x+\hat{\mu})\, S(x+\hat{\mu},y) \\
&\quad - U(x,x-\hat{\mu})\, S(x-\hat{\mu},y)\,] \\
&= \delta(x,y)
\end{aligned} \quad (5)$$

The use of an asymmetric forward time difference facilitates implementing the requirement that heavy quarks propagate only forward in time. A symmetrical difference could also be used, at the cost of a small increase in computation.

The free inverse propagator corresponding to the forward-time symmetrical-space discretization is

$$S^{(0)-1}(p) = v_0(e^{ip_4}-1) + \sum_{\mu=1}^{3} v_\mu \sin p_\mu \quad (6)$$

The balance of this talk will be devoted to a study of some rather unfamiliar structural aspects of the theory defined by this reduced discrete Dirac equation.

## 3. THE LATTICE HQET SPECTRUM

We will discuss together fermion doubling and the asymptotic behavior of the reduced heavy quark propagator. Doubling occurs because the symmetrical first spacial difference results in the presence of modes with large residual momentum but small residual energy. Growing modes occur because the reduced heavy quark Euclidean-space propagator is obtained from the full propagator by factoring out the zero residual momentum decay. This results in those modes whose residual momenta are opposite to the classical velocity growing with Euclidean time. In the heavy quark limit, the rate of decay of the mode with residual momentum $\vec{p}$ is given by the finite residual energy

$$\sqrt{M^2+(M\vec{v}+\vec{p})^2} - M\sqrt{1+\vec{v}^2} \to \frac{\vec{v}\cdot\vec{p}}{v_0} \quad (7)$$

When this is negative, the "falloff" becomes growth. Conversely, at fixed Euclidean time, the heavy quark propagator grows with momentum in directions opposite to $v$.

The growing modes do not lead to physical effects or loop divergences in the LHQET because they are always accompanied by modes of either a light Wilson quark or a gluon whose momentum is at the edge of the Brillouin zone. The suppression of divergences associated with the growing heavy quark modes is seen explicitly in perturbative loop diagrams.[6] It occurs because the rate of falloff of the modes of the light quantum is $\sqrt{|\vec{p}|^2+m^2}$, while according to Eq. (7), the rate of heavy quark propagator's growth is only a finite fraction of $|\vec{p}|$ (since $|\vec{v}|/v_0$ is strictly less than 1). In the Euclidean energy plane, these effects appear as a result of the integration contour, determined by the heavy quark's propagating only forward in time.

The absence of physical manifestations of the heavy quark doubler modes is illustrated by lattice free field theory.[7] The falloff of a reduced composite propagator for a "meson" made of a free heavy quark and a free light quark with momentum $P = M\vec{v}+\vec{p}$ is given by the minimum of the combined energies of a light quark with momentum $\vec{k}$ and a heavy quark with momentum $M\vec{v}+\vec{p}-\vec{k}$, reduced by the heavy quark energy for vanishing residual momentum, $Mv_0$. In two dimensions the minimization can be carried out analytically. The minimum is unique; only one heavy quark mode is selected. In the heavy quark limit, the minimum occurs at $k_1 = mv_1$, and is the correct residual energy for an object with mass $(M + m)$, classical velocity $v$, and residual momentum $p_1$, namely

$$\sqrt{(Mv_1 + p_1)^2 + (M + m)^2} - M v_0 \qquad (8)$$
$$\rightarrow (v_1/v_0) p_1 + m/v_0$$

## 4. RENORMALIZATION OF THE CLASSICAL VELOCITY

A new feature of the HQET on the lattice is the possibility that the classical velocity is renormalized. This is a finite lattice artifact directly related to the reduction of the spatial symmetry from $O(4)$ to the hypercubic group and can be estimated both in lattice perturbation theory and in simulations. The full inverse propagator for a heavy quark with classical velocity $v$ is given by

$$S^{-1}(p) = S^{(0)-1}(p) - \Sigma(p) \qquad (9)$$

where in perturbation theory $\Sigma$ is the sum of all proper self mass diagrams. Denoting the derivative of $\Sigma$ with respect to $p$ at zero by $X_\mu$, the renormalized classical velocity is given by

$$\left. \frac{\partial S_v^{-1}(p)}{\partial p_\mu} \right|_{p=0} = v_\mu^{(0)} - X_\mu \qquad (10)$$
$$= Z^{-1} v_\mu^{ren}$$

The requirement that the physical classical velocity be normalized to 1 fixes both the renormalized classical velocity and the wave function renormalization constant $Z$.

$$Z = \frac{1}{\sqrt{(v_\mu^{(0)} - X_\mu)^2}} \qquad (11)$$
$$v_\mu^{ren} = Z (v_\mu^{(0)} - X_\mu)$$

In the continuum, $O(4)$ invariance forces $X$ to be proportional to $v$, the only available 4-vector, and so its only effect is to shift $Z$ from 1. On the lattice, however, there are additional linearly independent 4-vectors, such as $v_\mu^3$, which have exactly the same lattice transformation properties as $v_\mu$. Such terms are present in even the simplest 1 loop contribution to $\Sigma$, and they cause the classical velocity to be shifted by a finite amount.

We may use the change in energy for small residual momenta (Eq. (7)) to define the physical (renormalized) classical velocity without resort to perturbation theory. Of course, the quark propagator is gauge dependent, and so it is not an appropriate quantity to directly simulate. However, the propagator of a gauge invariant composite heavy-light meson propagator has the same dependence on the classical velocity as its heavy quark component, and is suitable for direct simulation. The falloff of a heavy-light propagator is controlled by the location of the closest zero (in the residual energy plane) of the inverse propagator. From Eq. (1) or (6) this gives

$$\left( \frac{v_i}{v_0} \right)^{(phys)} = \left. \frac{\partial E(v^{(0)}, p)}{\partial p_i} \right|_{\vec{p}=0} \qquad (12)$$

We have estimated the magnitude of the classical velocity renormalization both perturbatively and by simulating Eq. (12). The perturbative calculation follows the method of Ref. 6, with the replacement of the backwards time derivative used there with a forward derivative. For the simulation, we used a preliminary subset of $24^3 \times 48$ lattices with $\beta = 6.1$ and $\kappa = .154$ generously provided by the Fermilab Collaboration.[8] In carrying out the simulation, it proved useful to develop the propagator in an expansion in powers of $\tilde{v}_i = v_i/v_0$.

$$M(t,\vec{p}) = \sum \tilde{v}_1^{m_1} \tilde{v}_2^{m_2} \tilde{v}_3^{m_3} M(t,\vec{p},\vec{m}) \qquad (13)$$

This is effectively a transverse hopping expansion, because computing the propagator $m_i$ units from the origin in the $i^{th}$ direction introduces $m_i$ factors of $\tilde{v}_i$. For fixed $\tilde{v}$ and $\vec{p}$, the asymptotic decay of the propagator is

$$M(t,\vec{p}) \sim C(\vec{p}) e^{-E(\vec{p})t} \qquad (14)$$

The normalization and energy are also developed in analogous series. Parity invariance implies that terms even (odd) in the $i^{th}$ momentum component come only from power series coefficients with even (odd) values of $m_i$. The logarithmic derivative of the asymptotic propagator with respect to the residual momentum at the origin identifies, through Eq. (12), the physical classical velocity as the coefficient of $t$. The statistical signal seems to deteriorate quickly as higher order expansion coefficients are considered, so we restricted our preliminary simulation analysis to first order in the bare classical velocity. This gives

$$\frac{\partial M(t,\vec{p},m_i=1)/\partial p_i \big|_{\vec{p}=0}}{M(t,\vec{p}=0,m_i=0)} \quad (15)$$

$$\sim -\frac{\partial E(\vec{p},m_i=1)}{\partial p_i}\bigg|_{\vec{p}=0} t + const$$

The momentum derivative was approximated by the symmetrical first difference, and the ratio of $m_i = 1$ to $m_i = 0$ propagators was fit out to 6 units of Euclidean time separation.

Both the simulation and perturbation theory indicate that the physical classical velocity is smaller than its input, or bare value. The comparison of the shifts in the classical velocity, using the mean-field tadpole improved value for the coupling[9] corresponding to $\beta = 6.1$ in the perturbative evaluation is as follows:

$$\left(\frac{v_i^{(ren)}}{v_i^{(bare)}}\right)_{1\ loop} = .72$$

$$\left(\frac{v_i^{(phys)}}{v_i^{(input)}}\right)_{simul} = .6 \pm .1$$

These values are quite consistent, but perhaps surprisingly far from unity.

## 5. CONCLUSIONS

We have discussed in this talk some characteristics of the lattice implementation of the heavy quark effective theory. By means of essentially kinematic arguments, we argued that neither the asymptotic growth of some modes of the heavy quark propagator, nor the naive appearance of heavy fermion doubler modes leads to physical consequences.

We also quantitatively examined the (finite) renormalization of the classical velocity. This renormalization is peculiar to the LHQET, and does not occur in the continuum. At the values of the lattice coupling at which QCD simulations are typically performed today, this seems, on the basis of this very preliminary analysis, to be a significant effect. We of course plan to refine this analysis in the near future. The renormalization of the classical velocity must be taken into account in any attempt to use the LHQET to compute the Isgur-Wise function or other classical-velocity-dependent quantities, strictly in the heavy quark limit, on the lattice. As one evident example, the classical velocity rescaling changes the inferred values of the slope of the Isgur-Wise function at the origin.